\documentclass[12pt]{article}
\usepackage{amssymb,amsmath,amsfonts}
\usepackage[latin1,applemac]{inputenc}
\usepackage[T1]{fontenc}
\usepackage{seceqn}

\title{On the Birman-Schwinger principle applied to $\sqrt{-\Delta + m^2} - m$}
 
\author {M. Maceda \\[8pt]
               Mathematisches Institut\\
               Ludwig-Maximilians Universität, 80333 Munich}

\date{}

\def\b#1{{\mathbb #1}}

\def\Dirac{{\raise0.09em\hbox{/}}\kern-0.69em D}

\def\be{\begin{equation}}
\def\ee{\end{equation}}
\def\benum{\begin{enumerate}}
\def\eenum{\end{enumerate}}
\def\bestar{\begin{equation*}}
\def\eestar{\end{equation*}}
\def\norm #1#2{{\left\| #1 \right\|_{#2}}}

\def\mbb#1{{\mathbb #1}}

\def\stk#1 #2{{\stackrel #1 #2}}

\begin{document} 

\maketitle

\begin{abstract}
  The condition for $E = 0$ to be an eigenvalue of the operator $\sqrt{-\Delta + m^2} -m + \lambda V$ is obtained through the use of the Birman-Schwinger principle. By setting $E=-\alpha^2$ and using the analyticity of the corresponding Birman-Schwinger kernel, the series development of $(\lambda^{-1})(\alpha)$ is obtained up to second order on $\alpha$.
\end{abstract}

\vfill

\newpage

\section{Introduction}

The Birman-Schwinger principle~\cite{Birman:1961,Schwinger:1961} has been extensively used in several branches of physics and mathematics. 
The original motivation for its implementation was the possibility of having eigenvalue counting functions for the Schrödinger equation. In this approach, the potential term is taken to be of the form $\lambda V$ with $\lambda$ being a constant and $V$ a potential satisfying a certain number of properties (e. g., $V$ is non-positive). The eigenvalue counting functions thus obtained will depend on $\lambda$, and in general, for a given operator $\cal O$, they can be defined by the following expression
$$
N(\lambda) = \{\mbox{number of } \mu_j \in \mbox{Spec}({\cal  O}), \mu_j < \lambda\}.
$$
Research on this class of functions was actually initiated by Weyl for the Laplace operator $\Delta$~\cite{Weyl:1911,Weyl:1912}, with an emphasis in the following years on the asymptotic expression for $N(\lambda)$ when $\lambda \to \infty$. More recently, the use of Microlocal Analysis has indeed allowed a more detailed study of asymptotics with some proposals to replace eigenvalue counting functions by spectral shift functions~\cite{Gesztesy:1999}. 
 
 Generalizations of the method itself have also been proposed to deal with certain physical situations arising in quantum mechanics~\cite{Chadan:1999} and modified versions have been formulated for example in phonon physics where perturbations on acoustic and electromagnetic waves can be described through the presence of exponentially localized eigenmodes of their corresponding operators~\cite{Figotin:1996,Figotin:1997}, and where an analogy with the appearance of bound states for the Schrödinger equation can also be established.

 In~\cite{Klaus:1980} the eigenvalue equation
\be
(-\Delta + \lambda V) \psi = E \psi
\label{schrod}
\ee
was studied through the use of the Birman-Schwinger principle and some conditions expressing the fact that the solutions to the Schrödinger equation are in ${\cal L}^2 (\mbb R^3) $ were obtained. A related study was also done for the Dirac operator in~\cite{Klaus:1985}. In both cases the potential $V$ were assumed to satisfy a certain set of properties. In this work we shall apply the same techniques to the case of the pseudo-differential operator $\sqrt{-\Delta + m^2} -m$, which was studied by Herbst in the presence of a Coulomb potential~\cite{Herbst:1977}. This operator had also been used in other contexts such as hadronic physics~\cite{'tHooft:1974hx,Colangelo:1990rv,Lucha:1994tc}. For a discussion of several aspects of this relativistic operator and its relation to the Dirac operator we refer to~\cite{LeYaouanc:1993uc,Helffer:1994,Lucha:1994tc,Lucha:1994jp}.

In the next section we mention some basic facts regarding the Birman-Schwinger principle. Section 3 deals with the analyticity of the Birman-Schwinger kernel for $\sqrt{\Delta + m^2} - m$ and the series expansion of $\lambda^{-1}$ up to second order on $\alpha$. The Fourier transform of the coefficient proportional to $\alpha^2$ is then presented in Section 4. Several expressions used in the main text are derived in Appendix A and Appendix B. Appendix C contains a simple estimate on $|(\sqrt{-\Delta + m^2} - E)^{-1}(x)|$.

\section{The Birman-Schwinger principle}

  For convenience we recall the steps leading to the Birman-Schwinger principle and how it can be used  in order to extract some information about eigenvalues. One begins from a general eigenvalue equation of the form
\be
H \psi = (H_0 + \lambda V) \psi = E \psi
\label{hamilt}
\ee
with $H_0$ an operator representing the kinetic energy. We shall suppose in the following that the potential $V \in C_0^\infty (\b R^3)$ and takes negative values. The above equation then can be recast in the form
$$
\psi = \lambda (H_0 - E)^{-1} |V| \psi. 
$$
By multiplying this equation by $|V|^{1/2}$ and defining $\phi := |V|^{1/2} \psi$ one obtains 
\be
\lambda K \phi = \phi, \qquad K = |V|^{1/2} (H_0 - E)^{-1} |V|^{1/2}.
\label{birm-schw}
\ee
The physical meaning behind these rewriting is that the operator $K$ has an eigenvalue $\lambda_0^{-1}$ if and only if $H$ has an eigenvalue $E = E_0$ for $\lambda=\lambda_0$, the multiplicity in both cases being the same. One can think of this in some sense as transformation which preserves the multiplicity among the eigenvalues of these two different operators.

It is clear that the main feature of the method is to interchange the role played by the constant $\lambda$ and the energy $E$: the usual approach in quantum mechanics when dealing with an equation as Eq.~(\ref{schrod}) is to consider the potential term as being a perturbation and under suitable assumptions, the techniques of perturbation theory can then be employed to obtaine a (convergent) solution for the eigenfunctions $\psi$ and the eigenvalues $E$ in terms of the perturbation parameter $\lambda$. 
The Birman-Schwinger principle on the other hand allows one to consider the energy $E$ as a free parameter on which a series expansion for $\lambda$ can be obtained. In fact, for most practical reasons it is useful to set $E=-\alpha^2$ and consider a series development on powers of the parameter $\alpha$ (which is identified with the fine structure constant) as
\be
(\lambda^{-1})(\alpha)^{-1} = \lambda_0^{-1} + a\alpha + b\alpha^2 + \cdots.
\label{series}
\ee
After this is done, the question is to know if one can solve the above relation for $\alpha$ in terms of $\lambda$ and if the resulting series obtained in this way is analytic. As outlined in~\cite{Klaus:1980}, most of the steps leading to Eq.~(\ref{series}) pose no difficulties on this regard but a careful analysis has to be done for the second part.
 
\section{Kernel in coordinate space}

  In this section we study some properties of the Birman-Schwinger kernel for the operator $\sqrt{-\Delta + m^2} -m$. The integral kernel is given by Eqs.~(\ref{herbst}) and~(\ref{conv}) in Appendix A, namely
\be
K(x,y) = |V(x)|^{1/2} \frac m{4\pi |x - y|} \left[ \left( 1 - \frac {\mu^2}{m^2} \right)^{1/2} e^{-\mu |x - y|}  + \frac 2\pi F(m|x - y|; \mu)\right] |V(y)|^{1/2}.
\label{herbstk}
\ee
The parameter $\mu$ is determined from $\mu^2 = m^2 - (m + E)^2 = -2mE - E^2$. By setting $E = - \alpha^2$, one has $\mu^2 = 2m \alpha^2 - \alpha^4$; this will be used at the end of this section and in the next one.

In order to use perturbation theory on this operator, it is necessary to proof that it depends analytically on the parameter $\mu$. For convenience we shall set $m=1$ in the following. First, if one considers the case of real $\mu$, it is more or less clear that Eq.~(\ref{herbstk}) makes only sense as long as $0<\mu<1$, and thus one can try to look for analyticity inside the disc $|\mu|<1$ in the complex plane (we shall see that this is not so an arbitrary choice for the problem). Let us then consider $(f,Kg)$ for $f,g \in {\cal L}^2$. There are 4 different integrals over $\mbb R^3 \times \mbb R^3$ to be considered:
\begin{itemize}
\item $(1-\mu^2)^{1/2}(f, |V|^{1/2} \displaystyle \frac {e^{-\mu|\cdot|}}{|\cdot|}|V|^{1/2} g)$
\item $(f, |V|^{1/2} \displaystyle \frac {K_1(|\cdot|)}{|\cdot|}|V|^{1/2} g)$
\item $(1-\mu^2)(f, |V|^{1/2} e^{-\mu|\cdot|} \displaystyle \int_0^{|\cdot|} \cosh(\mu z) K_0(z) dz V|^{1/2} g)$
\item $(1-\mu^2)(f, |V|^{1/2} \sinh(\mu|\cdot|) \displaystyle \int_{|\cdot|}^\infty e^{-\mu z} K_0(z) dz |V|^{1/2} g)$
\end{itemize}
To deal with them we shall argue in the same way as in~\cite{Klaus:1985} for the Dirac operator. The argument is based on writing $(f,Kg)$ as a sum of integrals of the form
$$
c(\mu) \iint f(x) \frac {h(x,y)}{|x-y|^2} e^{-\mu|x-y|} g(y) d^3x d^3y 
$$
with $h(x,y)$ a function in ${\cal L}^\infty(\mbb R^3 \times \mbb R^3)$ having compact support and $c(\mu)$ an analytic function of $\mu$. By using Sobolev's inequality one then finds that $f(x) \frac {h(x,y)}{|x-y|^2} g(y) \in {\cal L}^1(\mbb R^3 \times \mbb R^3)$. The fact that $K$ as an operator can be bounded by operators with compact support in ${\cal L}^2$ means that one is dealing with a meaningful object. In this way the whole problem of an analytic extension on $\mu$ is contained in the expression of the coefficients $c(\mu)$ and the exponential $e^{-\mu|x|}$. 

For the first integral one immediately has $c(\mu) = (1-\mu^2)^{1/2}$ and $h(x,y)=|V(x)|^{1/2}|x-y||V(y)|^{1/2}$. In the second case obviously there is no dependence on $\mu$ and it suffices to note that with the function $h(x,y)=|V(x)|^{1/2}|x-y|K_1(|x-y|) |V(y)|^{1/2}$, which has compact support and is in ${\cal L}^\infty(\mbb R^3 \times \mbb R^3)$, one has $f(x) \frac {h(x,y)}{|x-y|^2} g(y) \in {\cal L}^1(\mbb R^3 \times \mbb R^3)$. 

The third and fourth integral require more care: first, for the third integral one has to consider $c(\mu) = (1-\mu^2)$, which is analytical on $\mu$, and a function 
$$
h_\mu(x,y) = |V(x)|^{1/2} |x-y| \int_0^{|x-y|} \cosh(\mu z) K_0(z) dz |V(y)|^{1/2}.
$$
However, it is clear that $h_\mu(x,y) < F_1(\mu) |V(x)|^{1/2} |x-y| |V(y)|^{1/2} $ where 
$$
F_1(\mu) = \int_0^\infty \cosh(\mu z) K_0(z) dz = \frac \pi{2 \sqrt{1-\mu^2}}
$$ 
as an elementary calculation shows. This also points to the necessity of considering $|\mu| \neq 1$.
One can now apply Sobolev inequality to the product $|V(x)|^{1/2} |x-y| |V(y)|^{1/2}$, obtaining thus that $f(x) \frac {h_\mu(x,y)}{|x-y|^2} g(y) \in {\cal L}^1(\mbb R^3 \times \mbb R^3)$. Furthermore, by writing $\cosh(\mu z)$ in terms of exponentials functions, it is clear that $h_\mu(x,y)$ is analytical on $\mu$.

Consider now the fourth integral: one has as before $c(\mu) = (1-\mu^2)$ and 
$$
h_\mu(x,y) = |V(x)|^{1/2} |x-y| \sinh(\mu |x-y|) \int_{|x-y|}^\infty e^{-\mu z} K_0(z) dz |V(y)|^{1/2}.
$$
One can now see that $h_\mu(x,y) < \frac 12 c |V(x)|^{1/2} |x-y| |V(y)|^{1/2} $, where $c=\int_0^\infty K_0(z) dz = \pi/2$, using the fact that $e^{-\mu z} \leq e^{-\mu a}$ for $z \in [a, \infty)$ and that $\int_a^\infty K_0(z) dz \leq \int_0^\infty K_0(z) dz$ for $a \geq 0$. Sobolev's inequality then leads as in the previous case to $f(x) \frac {h_\mu(x,y)}{|x-y|^2} g(y) \in {\cal L}^1(\mbb R^3 \times \mbb R^3)$. As before, the function $h_\mu$ is analytical on $\mu$ since it involves in a simple way analytical functions of $\mu$ in its definition.
From the previous considerations it follows then that $(f,Kg)$ is analytical on $\mu$ so that Theorem 3.12 in~\cite{Kato:1995} can be used to deduce that $K$ is analytic in the sense of norm. One should notice that the fact that one is considering a potential $V$ with compact support is fundamental for this argument to work.

Let us now proceed to finding under which conditions is 0 an eigenvalue of Eq.~(\ref{hamilt}). On general grounds one can argue in the following way:  given an operator $\cal O$ with inverse ${\cal O}^{-1}$, consider a function $\phi$ solution to the equation
\be
l_0 \phi = \mu_0 \phi
\ee
where $l_0 = |V|^{1/2} {\cal O}^{-1} |V|^{1/2}$. This equation is similar in structure to Eq.~(\ref{birm-schw}). Let us now define a function $u$ through the relation
$$
u = {\cal O}^{-1} |V|^{1/2} \phi.
$$
This function satisfies the integral equation
$$
u = {\cal O}^{-1} |V|^{1/2} \phi = \mu_0^{-1} {\cal O}^{-1} |V|^{1/2} |V|^{1/2} {\cal O}^{-1} |V|^{1/2} \phi = \mu_0^{-1} {\cal O}^{-1} |V| u, 
$$
namely
\be
u(x) = \mu_0^{-1} \int_{\mbox{supp } V} {\cal O}^{-1}(x, y) |V(y)| u(y) d^3 y. 
\ee
In the case of a potential $V$ with compact support, let us say $\Omega$, one obtains a development in series for large $|x|$ of the previous expression as
\be
u(x) = \mu_0^{-1} |x|^{-1} \int_\Omega c_1(y) |V(y)| u(y) d^3 y + \mu_0^{-1} |x|^{-2} \int_\Omega c_2(y) |V(y)| u(y) d^3 y + \dots
\label{powser}
\ee
and therefore
\be
\int_\Omega c_1(y) |V(y)| u(y) d^3 y = \mu_0 \int_\Omega c_1(y) |V(y)|^{1/2} \phi(y) d^3 y = 0
\label{cond1}
\ee
must be imposed in order to have $u \in {\cal L}^2(\b R^3)$. In the above the functions $c_1(y)$ and $c_2(y)$ are determined from the series development of ${\cal O}^{-1}(x, y)$ in powers of $1/|x|$ for large $|x|$. We have assumed that the negative powers of $|x|$ appearing in this series development are integers, which is clearly seen from Eq.~(\ref{powser}). This might be not necessarily the most general scenario, but as long as the first term has a factor $|x|^{-\gamma}$ with $\gamma \leq 3/2$, as in the case we are considering, Eq.~(\ref{cond1}) will be obtained. It can be pointed out that by using the relation $\mu_0^{-1} |V| u  = |V|^{1/2} \phi$, one has in the sense of distributions the following equation 
\be
({\cal O} - \mu_0^{-1} |V|) u = ({\cal O} + \mu_0^{-1} V) u = 0
\ee
and thus the condition stated in Eq.~(\ref{cond1}) is necessary and sufficient to guarantee that $0$ is an eigenvalue of the operator ${\cal O} + \mu_0^{-1} V$. 

We now go back to our case of study, ${\cal O} = \sqrt{-\Delta + m^2} - m$. We have from Eq.~(\ref{herbstk}) that when $E=0$,
\be
u(x) = \frac m{2\pi \mu_0} \frac 1{|x|} \int_\Omega |V(y)| u(y) d^3 y + \dots
\label{zcond}
\ee
for large $|x|$ since $c_1(y) = 1$ in Eq.~(\ref{cond1}). Therefore, according to the previous discussion, the constraint $\int |V(y)|^{1/2} \phi(y) d^3 y = 0$ is a necessary and sufficient condition for $0$ to be an eigenvalue. One should notice that a similar condition exists for the Schrödinger operator. This can be understood on the grounds of the non-local behavior associated to $\sqrt{-\Delta + m^2} - m$: non-locality is contained in the Bessel function $K_1(z)$ appearing in the function $F(m|x|; \mu = 0)$ in Eq.~(\ref{conv}) and for large distances (at least greater than $m^{-1}$) it becomes highly suppressed. In this situation the main contribution turns out to come from a Schrödinger-like term [first term in Eq.~(\ref{herbstk})]. 

One can also try to see what happens for small values of $|x|$. In this case there are two terms to lowest order to be considered,
\be
A_1 = \int_\Omega \frac 1{|y|} |V(y)| u(y) \, d^3 y
\ee
and 
\be
A_2 = \int_\Omega \frac 1{|y|} |V(y)| u(y) \int_{|y|}^\infty \frac{K_1(z)}z dz \, d^3 y.
\ee
Both terms are constants, independent of $|x|$. If $\mbox{supp }V = \Omega$ contains the origin, then for a non-divergent value of $u(|x| = 0)$ to exist, these constants should have finite values. The constant $A_1$ also appears when one considers a similar situation for the kernel of the Schrödinger operator. If the product $|V(y)| u(y)$ behaves like $y^{-\beta}$ with $\beta \leq 2$ near $0$ then it is well defined (strict inequality means $A_1$ = 0). On the other hand, $A_2$ is more singular due to the presence of the integral involving the Bessel function $K_1$. The first integral (from $|y|$ to infinity) diverges as $1/y$ near the origin. This means that finite values are possible in this case when the product $|V(y)| u(y)$ behaves like $y^{-\beta}$ with $\beta \leq 1$ (again, strict inequality means a vanishing value for $A_2$).

\section{Perturbation Theory and Fourier Transform}

In what follows we shall consider perturbation theory for Eq.~(\ref{herbstk}). From this kernel we can immediately write for $\alpha$ small the series development $K = L_0 + (2m)^{1/2} \alpha A + 2m \alpha^2 B + \dots$ where
\be 
L_0(x, y) = |V(x)|^{1/2} \frac m{4\pi |x - y|} \left[ 2 + \frac 2\pi \int_{m |x - y|}^\infty \frac {K_1(z)}z dz \right] |V(y)|^{1/2},
\label{zero-coeff}
\ee
\be
A(x, y) = - \frac m{2\pi} |V(x)|^{1/2} |V(y)|^{1/2},
\label{first-coeff}
\ee
and
\begin{multline}
B(x, y) = |V(x)|^{1/2} \frac 1{|x - y|} \left \{ \frac 12 (|x-y|^2 - \frac 1{m^2}) + \frac 1\pi (|x-y|^2 - \frac 2{m^2}) \int_0^{m|x-y|} K_0(z) dz  \right.  \\[4pt]
\left. + \frac {2|x - y|}{\pi m} \int_{m|x-y|}^\infty K_0(z) z dz + \frac 1{\pi m^2} \int_0^{m|x-y|} K_0(z) z^2 dz \right \} |V(y)|^{1/2}.
\label{second-coeff}
\end{multline}
As mentioned before, the presence of the second term inside the brackets in Eq.~(\ref{zero-coeff}) is an indication of the non-locality of the Herbst operator. Furthermore, by going into Fourier space, one can see that the kernel of $L_0$ is not positive definite, and for large values of $k$ it can take arbitrary positive and negatives values. This is due to the strong divergence produced by the integral of the quotient $K_1(z)/z$, which when coupled with the factor $1/|x-y|$ is of order 3.

Let us now find an expression for $\lambda^{-1} = (\phi, K \phi)$ [see Eq.~(\ref{birm-schw})] where $\phi$ satisfies $L_0 \phi = \mu_0 \phi$ and $\norm \phi2 = 1$. We have
\be
\begin{split}
(\lambda(\alpha))^{-1} =& (\phi, L_0 \phi) + (2m)^{1/2} \alpha (\phi, A \phi) + 2m \alpha^2 (\phi, B \phi)+ \dots  \\[4pt]
=& \mu_0  - \frac {\alpha m^{3/2}}{\sqrt 2 \pi} \left( \int_\Omega |V(y)|^{1/2} \phi(y) d^3 y \right)^2 + b \alpha^2 + \dots
\label{sker}
\end{split}
\ee
The expression for the coefficient $b$ is rather cumbersome as can be deduced from Eq.~(\ref{second-coeff}) and no definite statement on the sign of this coefficient can readily be established. But before with the analysis of it, let us review the general procedure used to write the energy $E$ as a function of the parameter $\lambda$. Assuming $\mu_0 \neq 0$, one deduces from the expression
$$
(\lambda(\alpha))^{-1} = \mu_0 + a \alpha + b \alpha^2 + \dots
$$ 
the relation
$$
\lambda_0^{-1} \lambda(\alpha) = 1 - \lambda_0 a \alpha + \lambda_0 (\lambda_0 a^2 - b) \alpha^2 + \dots, \qquad \lambda_0 := \mu_0^{-1},
$$
for $\alpha$ small. The next step is to invert this relation and to find $\alpha$ as a function of $\lambda$; it is here where care should be taken. If $a$ does not vanish then one has
$$
\alpha(\lambda) = - \frac 1{\lambda_0^2 a} (\lambda - \lambda_0) + \frac {\lambda_0 a^2 - b}{\lambda_0^4 a^3} (\lambda - \lambda_0)^2 + \dots
$$
and therefore 
$$
E(\lambda) = - \alpha(\lambda)^2 = - \frac 1{(\lambda_0^2 a)^2} (\lambda - \lambda_0)^2 + \dots
$$
On the other hand, if $a$ vanishes then
$$
\lambda_0^{-1} \lambda(\alpha) = 1 - \lambda_0 b \alpha^2 + c \alpha^3 + \dots
$$
and it follows that there are two different expressions for $\alpha$,  
$$
\alpha_{\pm}(\lambda) = ± \left[ \frac 1{\lambda_0^2 (-b)} (\lambda - \lambda_0) - \frac c{\lambda_0^3 (-b)^{3/2}} (\lambda - \lambda_0)^{3/2} + \dots \right]^{1/2}.
$$
One obtains then
$$
E_{\pm} (\lambda) = - \alpha_{\pm}(\lambda)^2 = - \frac 1{\lambda_0^2 (-b)} (\lambda - \lambda_0) + \dots
$$
and moreover, the series development is done in powers of $(\lambda - \lambda_0)^{1/2}$. 
For ${\cal O} = \sqrt{-\Delta + m^2} - m$, the coefficient $a$ is proportional to $\left( \displaystyle \int_\Omega |V(y)|^{1/2} \phi(y) d^3 y \right)^2$ according to Eq.~(\ref{sker}) and then it suffices to recall the discussion leading to Eq.~(\ref{zcond}) to conclude that the (first) second series development corresponds to $E=0$ (not) being an eigenvalue. For the first series, when $a \neq 0$, it is possible to appeal to the implicit theorem function to deduce analyticity of $\alpha(\lambda)$ at the point $\lambda_0$ and thus of $E(\lambda)$. When $a = 0$, analyticity does not hold in the Schrödinger case~\cite{Klaus:1980} because of an argument making use of the fact that $-\Delta + \lambda V$ only admits non-positive eigenvalues for all $\lambda>0$ (Theorem XIII.11 in~\cite{Reed:1978} applied to $V \in {\cal C}^\infty_0(\mbb R^3)$). Unfortunately we have been not able to find a similar result for the Herbst operator in the three-dimensional case. 

We also remark that in the case when $a=0$, if $\lambda$ approaches $\lambda_0$ from above by real values then $b$ should necessarily be negative for $\alpha_\pm(\lambda)$ to be well defined.
In the case of the Schrödinger equation it is known through an argument involving Fourier transform that $b$ takes negative values when $a$ vanishes~\cite{Klaus:1980}, in consequence the energy approaches a value $E_0=0$ from below as $\lambda \to \lambda_0^+$. For the Herbst operator one can proceed in the same way, i.e., one can write
$$
B(x,y) = |V(x)|^{1/2} {\cal B}(x,y) |V(y)|^{1/2}
$$
and therefore 
\be
b = 2m^2 (\phi, B \phi) = 2m^2(\phi, |V|^{1/2} {\cal B} |V|^{1/2}) = 2m^2(f, {\cal B} f) = 2m^2 \int \hat {\cal B}(k) |\hat f(k)|^2 d^3k
\label{b}
\ee
with $f = \phi |V|^{1/2}$. The sign of $b$ is then related to the behavior of the Fourier transform $\hat {\cal B}$. We should also note that the fact that $a=0$ guarantees that the expression in momentum space on the right hand side of Eq.~(\ref{b}) is defined as $k \to 0$.

As seen from Eq.~(\ref{second-coeff}) not all functions appearing in that expression have norm in ${\cal L}^2(\mbb R^3)$, and it is then necessary to use Fourier transform in the sense of distributions to find the Fourier transform $\hat{\cal B} (k)$. The details of this are given in Appendix B, here we only write the final result of the calculations which reads as
\begin{multline}
b = 2m^2 \int \hat {\cal B}(k) |\hat f(k)|^2 d^3k = 2 m \int (m^4\hat {\cal B}(m\sigma)) |\hat f(m\sigma)|^2 d^3\sigma \\[4pt]
=2 m \int \left[ - \frac 1{2\pi\sigma^2} -  \frac 1{4\pi^3\sigma^4} -\frac 1\pi \left[ \frac 1{8\pi^2\sigma^4} \frac{6w^4 + 5w^2 + 2}{(1 + w^2)^{5/2}} + \frac 1{\sigma^2}\frac 1{(1 + w^2)^{1/2}} \right]  \right. \\[4pt]
\left. + \frac 3{2\pi^2\sigma^3} \frac {w^3}{(1 + w^2)^{5/2}} - \frac 1{2\pi\sigma^2} \frac{2w^2 - 1}{(1 + w^2)^{5/2}} \right] |\hat f(m\sigma)|^2 d^3\sigma
\label{ftb}
\end{multline}
where $\vec \sigma = \vec k/m$, $\sigma = |\vec \sigma| \in [0, \infty)$ and $w = 2\pi\sigma$.
Even though a positive term is present in this expression, its contribution is dominated by similar negative terms as one can easily verify from Eq.~(\ref{ftb}). Therefore one concludes that $b$ is (conditionally) negative and that $E \to 0^-$ when $\lambda \to \lambda_0^+$.
Moreover, it also follows from this expression that in the non-relativistic case ($k = m\sigma \to 0$) the leading order contribution is given by $ -\pi^{-1} \sigma^{-2} - (2\pi^3)^{-1} \sigma^{-4}$, an analogous result to the Schrödinger case.

\section{Conclusions}
It has been shown that having wave-functions of the operator $\sqrt{-\Delta + m^2} - m$ in ${\cal L}^2(\mbb R^3)$ is closely related to the vanishing of the coefficient $a$ in the series development given by Eq.~(\ref{series}). This coefficient has the same form as in the Schödinger case as seen from Eq.~(\ref{sker}) and its vanishing can be recast as a statement on the fact that $E = 0$ is an eigenvalue of the equation $(\sqrt{-\Delta + m^2} - m + \lambda V) \phi = E \phi$. Moreover, the coefficient $b$ in Eq.~(\ref{series}) can be shown to consist of a Schrödinger-like part and more involved contributions due to the non-local nature of the operator under study. Fourier analysis however can be used to show that $b$ is conditionally negative. The physical consequence of these facts is similar to the Schrödinger case, namely, the phenomenon of "coupling constant threshold". 

When considering the Schrödinger operator with a spherically symmetric potential $V$ the above characterization can be explicitly formulated as a condition for the presence of $s$-waves as solutions of the equation $(-\Delta + V) \Psi = 0$ and indeed, it is only the behavior of $\Psi$ at infinity, where the potential $V$ vanishes, which should be considered to obtain this result~\cite{Klaus:1980}. 

In the case of the Herbst operator a similar criteria should be also possible, however there is a simple feature that is present in one case and not in the other: the answer for the Schrödinger operator can readily be given because one knows precisely the expression for $\Delta$ in spherical coordinates meanwhile for the Herbst operator one should study instead an integral equation. The following auxiliary problem then arises: is it possible to write an appropriate integral equation for $\Delta \Psi = 0$ from which one can recover the well known radial part $r^l, r^{-l-1}, \, l=0, 1, \dots$, of the spherically symmetric solutions? The answer is yes. The next natural step is to see if the same process can be applied to the integral equation obtained from $(\sqrt{-\Delta + m^2} - m)\Psi = 0$. This is a subject that we hope to discuss in a future work.

\section{Acknowledgements}
The author is grateful to E. Stockmeyer for having proposed the study of this problem as well as for remarks and discussions leading to the improvement of the text. Thanks are also due to S. Morozov for several discussions along this work. It is also recognized the financial support received through a grant from the Quantum and Analysis network, Contract No. HPRN-CT-2002-00277.

\appendix

\section*{Appendix A: Inverse Fourier transform}
\setcounter{section}{1}
\setcounter{equation}{0}

In this section we would like to derive the expression for $(\sqrt{-\Delta + m^2} - E)^{-1}$ in coordinate space. Since one is lead to deal with tempered function, the Fourier transform in the generalized sense should be considered. Several ways exist to calculate the Fourier transform of tempered functions, such as the $\epsilon$-prescription, where it is customary to consider after the calculations only those terms with no dependence on $\epsilon $ in the limit $\epsilon \to 0$. This method was widely used in the evaluation of propagators in the early days of QED. The Hankel transform~\cite{Bochner:1932} is other method introduced to deal with tempered functions. It is known that the $\epsilon$-prescription, the same being the case with the Hankel transform, is related to the notion of finite part, {\it Pf.} of a distributional pseudo-function~\cite[pp 20]{Schwartz:1955} acting on Schwartz's space $\cal S$ of infinitely differentiable and rapidly vanishing functions, together with its derivatives, at infinity. We shall also use the fact that for a distribution $f$, the relations $<{\cal F}(f), \phi> = <f, {\cal F}(\phi)>$ and $<{\cal F}^{-1}(f), \phi> = <f, {\cal F}^{-1}(\phi)>$ where $\phi \in \cal S$ hold~\cite{Bremermann:1965}.

Accordingly one can write
\bestar
\begin{split}
\int_{\mbb R^3} \frac 1{\sqrt{4\pi^2 p^2 + m^2}-E} \phi(p) d^3p =& E \int_{\mbb R^3} \frac 1{4\pi^2 p^2 + m^2 - E^2} \phi(p) d^3 p + \int_{\mbb R^3} \frac {\sqrt{4\pi^2 p^2 + m^2}}{4\pi^2 p^2 + m^2 - E^2} \phi(p) d^3 p  \\[4pt]
=& E \int_{\mbb R^3} \frac 1{4\pi^2 p^2 + m^2 - E^2} \phi(p) d^3 p  \\[4pt]
&+ (-\Delta + m^2) \int_{\mbb R^3} \frac 1{\sqrt{4\pi^2 p^2 + m^2}[4\pi^2 p^2 + \mu^2]} \phi(p) d^3p  \\[4pt]
=& E \int_{\mbb R^3} \frac 1{4\pi^2 p^2 + \mu^2} \phi(p) d^3 p \\[4pt]
&+ (-\Delta + \mu^2)  \int_{\mbb R^3} \frac 1{\sqrt{4\pi^2 p^2 + m^2}[4\pi^2 p^2 + \mu^2]} \phi(p) d^3p  \\[4pt]
&+ (m^2 - \mu^2)  \int_{\mbb R^3} \frac 1{\sqrt{4\pi^2 p^2 + m^2}[4\pi^2 p^2 + \mu^2]} \phi(p) d^3p.
\end{split}
\eestar
where we have set $\mu^2 = m^2 - E^2$. Therefore, from $<{\cal F}(f), \phi> = <f, {\cal F}(\phi)> = <{\cal F}^{-1}({\cal F}(f)), \phi>$, one has in the sense of distributions acting on $\cal S$, 
\begin{multline*}
{\cal F}^{-1}\left[ \frac 1{\sqrt{4\pi^2 p^2 + m^2}-E} \right] (x) =  E {\cal F}^{-1}\left[ \frac 1{4\pi^2 p^2 + \mu^2}\right](x) \\[4pt]
+ (-\Delta + \mu^2) {\cal F}^{-1}\left[ \frac 1{\sqrt{4\pi^2 p^2 + m^2}[4\pi^2 p^2 + \mu^2]} \right](x)  \\[4pt]
+ (m^2 - \mu^2) {\cal F}^{-1}\left[ \frac 1{\sqrt{4\pi^2 p^2 + m^2}[4\pi^2 p^2 + \mu^2]} \right](x) .
\end{multline*}

The first term is a standard example in textbooks on Fourier transforms or potential theory and can be easily found: 
\bestar
\begin{split}
\int_{\mbb R^3} \frac 1{4\pi^2 p^2 + \mu^2} \phi(p) d^3 p &= \int_{\mbb R^3} \int_0^\infty e^{-(4\pi^2 p^2 + \mu^2)t} dt \phi(p)d^3p \\[4pt]
&= \int_{\mbb R^3} \int_0^\infty \int_{\mbb R^3} e^{-4\pi^2 p^2t + 2\pi p\cdot x} d^3p e^{-\mu^2 t} {\cal F}(\phi)(x) d^3 x  \\[4pt]
&= \int_{\mbb R^3} \int_0^\infty \frac 1{(4\pi t)^{3/2}} \exp[ - |x|^2/4t - \mu^2 t] dt {\cal F}(\phi)(x) d^3 x  \\[4pt]
&= \frac {\mu}{(4\pi)^{3/2}} 2  \sqrt{\frac 2{\mu|x|}} \int_{\mbb R^3} K_{1/2}(m|x|) {\cal F}(\phi)(x) d^3 x  \\[4pt]
&= \int_{\mbb R^3} \frac {e^{-\mu |x|}}{4\pi|x|} {\cal F}(\phi)(x) d^3 x
\end{split}
\eestar
with $\phi \in \cal S$. In the above we have use Fubini's theorem and an integral representation of the Bessel functions $K_\nu(z)$. 
For the second and third term one should calculate the inverse Fourier transform
$$
{\cal F}^{-1}\left[ \frac 1{\sqrt{4\pi^2 p^2 + m^2}[4\pi^2 p^2 + \mu^2]} \right](x).
$$
Using the fact that the usual properties of convolution of the (inverse) Fourier transform remain valid for distributions~\cite{Bremermann:1965}, then the problem reduces to the convolution of ${\cal F}^{-1}\left[ \frac 1{4\pi^2 p^2 + \mu^2} \right] (x)$ with ${\cal F}^{-1}\left[ \frac 1{\sqrt{4\pi^2 p^2 + m^2}} \right] (x)$. This latter inverse Fourier transform can be found in the same way as done before with the result
$$
\int_{\mbb R^3} \frac 1{\sqrt{4\pi^2 p^2 + m^2}} \phi(p) d^3 p = \frac m{2\pi^2} \int_{\mbb R^3} \frac {K_1(m |x|)}{|x|} {\cal F}(\phi)(x) d^3 x.
$$
Therefore, in distributional sense, one has the following expression
\bestar
\begin{split}
{\cal F}^{-1}\left[ \frac 1{\sqrt{4\pi^2 p^2 + m^2}-E} \right] (x) =& \frac E{4\pi |x|} e^{-\mu|x|} + (-\Delta + \mu^2) \left[ \frac {e^{-\mu|\cdot|}}{4\pi |\cdot|} \ast \frac m{2\pi^2} \frac{K_1(m|\cdot|)}{|\cdot|} \right](x)  \\[4pt]
&+ \frac {m(m^2 - \mu^2)}{8\pi^3} \left[ \frac {e^{-\mu|\cdot|}}{|\cdot|} \ast \frac{K_1(m|\cdot|)}{|\cdot|} \right](x)  \\[4pt]
=& \frac E{4\pi |x|} e^{-\mu|x|} + \frac m{2\pi^2} \frac{K_1(m|x|)}{|x|}  \\[4pt]
&+ \frac {m(m^2 - \mu^2)}{8\pi^3} \left[ \frac {e^{-\mu|\cdot|}}{|\cdot|} \ast \frac{K_1(m|\cdot|)}{|\cdot|} \right](x)
\end{split}
\eestar
It is clear from the above that this expression might be treated indeed as an ordinary function if the convolution in the last line makes sense. In fact, using the expansion~\cite{Abramowitz:1965}
$$
\frac {e^{-\lambda R}}{\lambda R} = \frac 2\pi \sum_{l=0}^\infty (2l + 1) \left[ \sqrt{\frac \pi{2\lambda r_1}} I_{l + \frac 12} (\lambda r_1) \right] \left[ \sqrt{\frac \pi{2\lambda r_2}} K_{l + \frac 12} (\lambda r_2) \right] P_l (\cos \gamma)
$$
where $R = [r^2 + \rho^2 - 2r \rho \cos(\gamma)]^{1/2} , r_1 = \mbox{min}(r, \rho), r_2 = \mbox{max}(r, \rho)$, only the term with $l=0$ is seem to give a non-vanishing contribution for the convolution after performing the integral over the angular variables. Therefore, one obtains
\bestar
\begin{split}
\left[ \frac {e^{-\mu|\cdot|}}{|\cdot|} \ast \frac{K_1(m|\cdot|)}{|\cdot|} \right](x) =& \frac {4\pi}{m^2} \frac {2\mu}\pi \sqrt{\frac \pi{2\mu |x|}} K_{\frac 12} (\mu |x|) \int_0^{m |x|} \sqrt{\frac {m\pi}{2\mu y}} I_{\frac 12} (\frac {\mu y}m) K_1(y) y dy  \\[4pt]
&+ \frac {4\pi}{m^2} \frac {2\mu}\pi \sqrt{\frac \pi{2\mu |x|}} I_{\frac 12} (\mu |x|) \int_{m |x|}^\infty \sqrt{\frac {m\pi}{2\mu y}} K_{\frac 12} (\frac {\mu y}m) K_1(y) y dy  \\[4pt]
=& \frac {4\pi}{m^2 |x|} e^{-\mu |x|} \int_0^{m |x|} \frac m{\mu y} \sinh\left( \frac {\mu y} m \right) K_1(y) y dy  \\[4pt]
&+ \frac {4\pi}m \frac {\sinh(\mu |x|)}{\mu |x|} \int_{m |x|}^\infty e^{-\mu y/m} K_1(y) dy.
\end{split}
\eestar 
The explicit expressions of the Bessel functions $I_{1/2}(x)$ and $K_{1/2}(x)$ have been used in order to write the above relation. The final outcome of the previous calculations is then
\be
\begin{split}
{\cal F}^{-1}\left[ \frac 1{\sqrt{4\pi^2 p^2 + m^2}-E} \right] (x) =& \frac m{4\pi |x|} \left[ \frac Em e^{-\mu |x|}  + \frac 2\pi F(m|x|; \mu)\right]  \\[4pt]
=& \frac m{4\pi |x|} \left[ \left( 1 - \frac {\mu^2}{m^2} \right)^{1/2} e^{-\mu |x|}  + \frac 2\pi F(m|x|; \mu)\right]
\label{herbst}
\end{split}
\ee
with
\be
\begin{split}
F(m|x|; \mu) =& K_1(m|x|) + \frac {m^2 - \mu^2}{4\pi} |x| \left[ \frac {e^{-\mu|\cdot|}}{|\cdot|} \ast \frac{K_1(m|\cdot|)}{|\cdot|} \right] (x) \\[4pt]
=& K_1(m|x|) + \left( 1 - \frac {\mu^2}{m^2} \right) \left[ e^{-\mu|x|} \int_0^{m|x|} \left[ \frac m{\mu y} \right] \sinh\left( \frac {\mu y}m \right) K_1(y) y dy \right.  \\[4pt]
&\left. + \frac m\mu \sinh( \mu |x| )  \int_{m|x|}^\infty  e^{-\mu y/m} K_1(y) dy \right]  \\[4pt]
=& K_1(m|x|) + \left( 1 - \frac {\mu^2}{m^2} \right) \left[ e^{-\mu|x|} \int_0^{m|x|} \cosh\left( \frac {\mu y}m \right) K_0(y) dy \right.  \\[4pt]
&\left. - \sinh\left( \mu|x| \right) \int_{m|x|}^\infty e^{-\mu y/m} K_0(y) dy \right]. 
\label{conv}
\end{split}
\ee
In the above we have used some relations among integrals involving Bessel and exponentials functions~\cite{Luke:1962,Abramowitz:1965} in order to write the last expression.

\section*{Appendix B: Fourier transform of ${\cal B} (x,y)$}
\setcounter{section}{2}
\setcounter{equation}{0}

For convenience we present here the calculations leading to Eq.~(\ref{ftb}) in the text. In the expression for $\cal B$, several tempered functions in the sense of distributions appear and their Fourier transforms, which will be also tempered functions, can be found in this context. 

In the following we shall apply the Hankel transform to a well-behaved function suited to the problem (strictly speaking the Hankel transform refers to the one-dimensional case) and then argue its validity for more general cases.

To begin with we shall use the formula~\cite[\S 43]{Bochner:1932}
$$
{\cal F}(f,k) = \frac {2\pi}{k^{\frac {n-2}2}} \int_0^{+\infty} dr f(r) r^{n/2} J_{\frac {n-2}2}(2\pi kr).
$$
Here $n$ is the dimension of the space considered. Let us apply this to the function $f(x) = |x|^{-\alpha} \int_0^{m|x|} g(z) dz$ with $0 < \alpha$, we have 
\bestar
\begin{split}
{\cal F}(f,r) =& \frac {2\pi}{r^{\frac {n-2}2}} \int_0^{\infty} dr  r^{-\alpha} \int_0^{mr} dz g(z) r^{n/2} J_{\frac {n-2}2}(2\pi kr)  \\[4pt]
=& \frac {2\pi}{r^{\frac {n-2}2}} \int_0^{\infty} dr r^{n/2 - \alpha}  J_{\frac {n-2}2}(2\pi kr)  \int_0^{mr} dz g(z)  \\[4pt]
=& \frac {2\pi}{k^{\frac {n-2}2}} \frac 1{(2\pi k)^{n/2 - \alpha + 1}} \int_0^{\infty} dz g(z) \int_{z/m}^\infty du u^{n/2 - \alpha}  J_{\frac {n-2}2}(u) \\[4pt]
=& \frac 1{(2\pi)^{n/2-\alpha}} \frac 1{k^{n-\alpha}} \left[ \int_0^{\infty} dz g(z) \int_0^\infty du u^{n/2-\alpha} J_{\frac {n-2}2}(u) \right. \\[4pt] 
&\left. - \int_0^{\infty} dz g(z) \int_0^{wz} du u^{n/2-\alpha} J_{\frac {n-2}2}(u) \right]   
\end{split}
\eestar
In the above we have defined a dimensionless parameter $w = 2\pi k/m$ and assumed that the function $g(z)$ is well behaved so that Fubini's theorem can be used.
For our function $g(z)$ we shall take indeed $g(z) = z^\beta K_0(z), 0 \leq \beta$. The previous expression is justified as long as $n/2 < \alpha < n$ since the integrals involving the Bessel function $J_\nu(u)$ are then well defined. After using the relations
\bestar
\begin{split}
\int_0^\infty s^\mu J_\nu (s) ds &= 2^\mu \Gamma(\frac {\nu + \mu + 1}2)/\Gamma(\frac {\nu - \mu + 1}2),  \\[4pt]
\int_0^\infty s^\mu K_\nu (s) ds &= 2^{\mu-1} \Gamma(\frac {\mu + \nu + 1}2) \Gamma(\frac {\mu - \nu + 1}2)
\end{split}
\eestar
and
$$
\int_0^x s^\mu J_\nu (s) ds = x^{\mu + 1} \sum_{k=0}^\infty \frac {(-1)^k (x/2)^{\nu + 2k}}{k! (\mu+\nu+2k+1)\Gamma(\nu+k+1)}
$$
one arrives to
\bestar
\begin{split}
{\cal F}(f,r) =& \frac 1{(2\pi)^{n/2-\alpha}} \frac 1{k^{n-\alpha}} \left[ 2^{\beta + n/2 - \alpha -1} \Gamma\left(\frac {\beta+1}2\right)^2 \frac {\Gamma(\frac {n-\alpha}2)}{\Gamma(\frac \alpha2)} \right. \\[4pt]
&\left. - w^{n-\alpha} 2^{\beta + n/2 - \alpha -1}  \sum_{m=0}^\infty \frac {(-w^2)^m}{m! (m + \frac {n-\alpha}2)} \frac {\Gamma(m+\frac {\beta + n -\alpha +1}2)^2}{\Gamma(m + \frac n2)} \right].
\end{split}
\eestar
The sum inside the brackets is just a hypergeometric function; the final result then reads as
\be
\begin{split}
{\cal F}(f,r) =& \frac 1{(2\pi)^{n/2-\alpha}} \frac 1{k^{n-\alpha}} \left[ 2^{\beta + n/2 - \alpha -1} \Gamma\left(\frac {\beta+1}2\right)^2 \frac {\Gamma(\frac {n-\alpha}2)}{\Gamma(\frac \alpha2)} 
- w^{n-\alpha} 2^{\beta + n/2 - \alpha}  \frac {\Gamma(\frac {\beta+ n-\alpha + 1}2)^2}{(n - \alpha)\Gamma(\frac n2)} \right. \\[4pt]
&\left. \times {}_3F_2(\frac {n-\alpha}2, \frac {\beta+ n-\alpha + 1}2, \frac {\beta+ n-\alpha + 1}2; \frac n2, 1 + \frac {n-\alpha}2; -w^2)\right].
\label{ft1}
\end{split}
\ee
Using a reasoning due to L. Schwartz~\cite[pp 113]{Schwartz:1957}, this expression can be seen to hold also for other values of $\alpha$, in particular for $\alpha < 0$, if one keeps in mind that the function $f$ should be then understood as a tempered function as well as its Fourier transform. More care should be taken when $n-\alpha = -2h$, $\alpha = -2h$ or $\beta + n-\alpha +1 = -2h$, $h$ being a positive integer, where poles appear in the $\Gamma$ function, but the appropriate modification to be used for those cases has also been pointed out by Schwartz. This modification will not concern us however because of the values of $\alpha$ ($\pm 1$ or $0$) appearing in Eq.~(\ref{second-coeff}) and of the dimension $n=3$. 
As a particular case then one can consider $\alpha=1, \beta = 0, n=3$,
\bestar
\begin{split}
{\cal F}(f,r) &= \frac 1{(2\pi)^{1/2}} \frac 1{k^2} \left[ \frac {\pi^{1/2}}{2^{1/2}} - \frac {w^2}{2^{1/2}} \Gamma(3/2) {}_3F_2(1,3/2,3/2; 3/2, 2; -w^2) \right]  \\[4pt]
&= \frac 1{2 k^2} \frac 1{(1+ w^2)^{1/2}}.
\end{split}
\eestar
This corresponds, taking into account a factor $-2/(\pi m^2)$ and replacement $k = m\sigma$, to the second term inside the brackets in Eq.~(\ref{ftb}).  

It is also straightforward from the previous computation to identify the Fourier transform of $f(x) = |x|^{-\alpha} \displaystyle \int_{m|x|}^\infty g(z) dz, 0 < \alpha$, this is given by the second term in Eq.~(\ref{ft1}), namely
\be
\begin{split}
{\cal F}(f,r) =& \frac 1{(2\pi)^{n/2-\alpha}} \frac 1{k^{n-\alpha}} w^{n-\alpha} 2^{\beta + n/2 - \alpha} \frac {\Gamma\left(\frac {\beta+ n-\alpha+1}2\right)^2}{(n-\alpha)\Gamma(\frac n2)} \\[4pt]
&\left. \times {}_3F_2(\frac {n-\alpha}2, \frac {\beta+ n-\alpha + 1}2, \frac {\beta+ n-\alpha + 1}2; \frac n2, 1 + \frac {n-\alpha}2; -w^2)\right].
\end{split}
\ee
This formula can also be generalized to other values of $\alpha$ than those in the interval $(n/2, n)$, care needed only to be taken in more detail at the points mentioned in a previous paragraph. Taking this into account, if we now calculate the case $\alpha=0, \beta=1, n=3$, the following expression is obtained
\bestar
\begin{split}
{\cal F}(f,r) &= \frac 1{(2\pi)^{3/2}} \frac 1{k^3} w^3 2^{5/2} \frac {\Gamma(5/2)^2}{3\Gamma(3/2)} {}_3F_2(3/2,5/2,5/2; 3/2,5/2; -w^2)  \\[4pt]
&= \frac 3{4\pi} \frac 1{k^3} w^3 2^{3/2} \frac 1{(1+w^2)^{5/2}}
\end{split}
\eestar
corresponding, times a factor $2/(\pi m)$ and replacement $k=m\sigma$, to the positive term in Eq.~(\ref{ftb}).

\section*{Appendix C: An estimate}
\setcounter{section}{3}
\setcounter{equation}{0}

We rewrite Eq.~(\ref{herbst}) as
\bestar
(\sqrt{ - \Delta + m^2} - E)^{-1} (|x|) = \frac m{4\pi |x|^2} \left[ \left( 1 - \frac {\mu^2}{m^2} \right)^{1/2} |x| e^{-\mu |x|}  + \frac 2\pi |x| F(m|x|; \mu)\right].  
\eestar
We would like to show that for a fixed value of $\mu$, each term inside the brackets is bounded as function of $|x|$. This is obviously true for $h_0(|x|) = |x| e^{-\mu |x|}$ ($\leq \mu^{-1}$) and also for the first term coming from the product $|x| F(m|x|; \mu)$, namely $h_1(|x|) = |x| K_1(m|x|)$ ($\leq 1$). The next two terms to consider are
$$
h_2(|x|) = |x| e^{-\mu |x|} \int_0^{m|x|} \cosh\left( \frac {\mu y}m \right) K_0(y) dy
$$
and
$$
h_3(|x|) = |x| \sinh\left( \mu|x| \right) \int_{m|x|}^\infty e^{-\mu y/m} K_0(y) dy.
$$
$h_2$ can be seen to be bounded as follows
\begin{multline*}
h_2(|x|) \leq \mu^{-1} \int_0^\infty \cosh\left( \frac {\mu y}m \right) K_0(y) dy = \frac 1{2\mu} \int_0^\infty \cosh\left( \frac {\mu y}m \right) \int_0^\infty \exp \left[ - \frac 1u - \frac {u y^2}4 \right] \frac {du}u dy  \\[4pt]
= \frac 1{4\mu} \int_0^\infty \int_{-\infty}^\infty \exp \left[ - \frac {u y^2 }4 + \frac {\mu y}m \right] dy \, e^{-1/u} \frac {du}u = \frac {\pi^{1/2}}{2\mu} \int_0^\infty \exp \left[ -\frac 1u \left( 1 - \frac{\mu^2}{m^2} \right) \right] \frac {du}{u^{3/2}}  \\[4pt]
= \frac \pi{2\mu} \left( 1 - \frac{\mu^2}{m^2} \right)^{-1/2}.
\end{multline*}
Let us turn now to $h_3$, one has
\bestar
\begin{split}
h_3(|x|) & \leq |x| e^{\mu |x|} \int_{m|x|}^\infty K_0(y) e^{-\mu y/m} dy \leq |x| \int_{m|x|}^\infty K_0(y) dy  \\[4pt]
& \leq m|x_0|^2 K_0(m|x_0|)
\end{split}
\eestar
where $m|x_0|$ satisfies the transcendental equation $\int_z^\infty K_0(y) dy = z K_0(z)$ with solution $z=0.7451315$. From this follows also that $m|x_0| K_0(m|x_0|) < \pi/2$ and hence
$$
h_3(|x|) < \pi |x_0| / 2 \leq c\pi / 2m, \qquad c \in [0.7451315, \infty).
$$
Collecting all the previous results we have the following estimate
\begin{multline}
| (\sqrt{ - \Delta + m^2} - E)^{-1} (|x|) | \leq \frac m{4\pi |x|^2} \left[ \left( 1 - \frac {\mu^2}{m^2} \right)^{1/2} \frac 1\mu \right.  \\[4pt] 
+ \left. \frac 2\pi \left[ 1 + \left( 1 - \frac {\mu^2}{m^2} \right) \left[ \frac \pi{2\mu} \left( 1 - \frac{\mu^2}{m^2} \right)^{-1/2} + \frac {c\pi}{2m} \right]  \right] \right]  \\[4pt]
\leq \frac m{4\pi |x|^2} \left[ 1 + \frac 2\mu + \frac cm \right].
\end{multline}

\newpage

\setlength{\parskip}{5pt}

\providecommand{\href}[2]{#2}\begingroup\raggedright\endgroup

\end{document}